# Application Embedding: A Language Approach to Declarative Web Programming

David H. Lorenz[a,b*] and Boaz Rosenan[c]

a Open University of Israel

b Faculty of Computer Science, Technion—Israel Institute of Technology

c University of Haifa

**Abstract** Since the early days of the Web, web application developers have aspired to develop much of their applications declaratively. However, one aspect of the application, namely its business-logic is constantly left imperative. In this work we present *Application Embedding*, a novel approach to application development which allows all aspects of an application, including its business-logic, to be programmed declaratively.

We develop this approach in a two-step process. First, we draw a mapping between web applications and *Domain-Specific Languages (DSLs)*. Second, we note that out of the two methods for implementing DSLs, namely as either *internal* or *external*, most traditional web applications correspond to external DSLs, while the the technique that corresponds to DSL embedding (implementing internal DSLs) is left mostly unexplored.

By projecting the well-known technique of DSL embedding onto web applications, we derive a novel technique—*Application Embedding*. Application embedding offers a separation of code assets that encourages reuse of imperative code, while keeping all application-specific assets, including those specifying its business-logic, declarative.

As validation, we implemented a simple, though nontrivial web application using the proposed separation of assets. This implementation includes an application-agnostic imperative host application named `FishTank`, intended to be applicable for a wide variety of web applications, and a declarative definition of the different aspects of the specific application, intended to be loaded on that host.

Our method of separation of code assets facilitates a better separation of work, in comparison to traditional methods. By this separation, host application developers can focus mostly on the extra-functional aspects of a web application, namely on improving performance, scalability, and availability, while developers of an embedded application can focus on the functional aspects of their application, without worrying about extra-functional concerns. The reusability of the host application makes the effort put into a better implementation cost-effective, since it can benefit all applications built on top of it.

# 1 Introduction

Web application development has gone a long way to make much of the application code declarative. This includes the introduction of server-side frameworks (e.g., Ruby on Rails[1] and Django[2]), client-side frameworks (e.g., AngularJS[3] and React[4]), and languages backed by web standards (e.g., HTML and CSS). While these frameworks and languages make declarative programming applicable to many of these applications' concerns (most notably, the presentation), one important aspect of web applications is constantly left imperative: their business logic.[5]

In this paper we apply a well-known principle in Computer Science—the notion that *program* (code) and *state* (data) are interchangeable [38, 4]—to another well known technique—*domain-specific language embedding* [23, 21]—to form a new approach to application development, named *Application Embedding*, allowing application business logic to be *defined declaratively* rather than *implemented imperatively*.

Applying declarative programming, which is stateless by nature, to business logic implementation, which is imperative by nature, may seem at first sight implausible or even impossible. Indeed, imperative, general-purpose programming languages are often used to implement business logic. However, if we take a closer look, imperative business logic can be attributed to a common but inessential design choice: the *Three-Tier Architecture (3TA)* [13].

## 1.1 Traditional Approach

In the 3TA a web application is partitioned into a presentation-tier running the presentation logic on the client-side, a logic-tier running the business logic typically on a web server on the sever-side, and a data-tier consisting of a database program also running on the server-side.

The 3TA provides a separation of concerns that promotes *scalability*. Stateless components are scaled easily by adding more computer nodes running them. Stateful components, however, are harder to scale. For this reason, the 3TA segregates the latter parts in their own tier (namely, the data-tier), requiring them to be application agnostic (e.g., a general-purpose database), so once implemented, they can be reused massively across applications, thus amortizing the cost of their development.

---

[*] Work done in part while visiting the Faculty of Computer Science, Technion—Israel Institute of Technology.

[1] http://rubyonrails.org/

[2] https://www.djangoproject.com/

[3] https://angularjs.org/

[4] https://facebook.github.io/react/

[5] The term *business logic* in web applications refers to the part of the application that manipulates the data (state) that the application stores, and it is used in a broader sense than its original meaning in enterprise applications where it refers strictly to business rules or business processes. For example, a decision in a social network regarding the placement of a user status in another user's timeline is a part of the social network's business logic, although this decision has nothing to do with business *per se*.

Consequently, however, the 3TA separates computation and state into two separate entities (the logic-tier and the data-tier, respectively). This means that the logic-tier needs to perform I/O operations to interact with the application's state, forcing its imperative nature. Some web frameworks and multi-tier programming languages strives to hide away the imperative nature of the logic-tier, thus allowing developers to define most of the application's concerns declaratively. However, they do so typically by relying on the declarative nature of relational databases, thus assuming the application's data model maps nicely to a normalized relational database. Data models that are hard to fit into relational data (e.g., due to impedance mismatch [35]), or large systems that require denormalization to meet performance goals, break these assumptions and require imperative code to fill the gap.

### 1.2 A Domain-Specific Language Approach

We present a different separation of concerns. Our method of partitioning keeps all application-specific parts declarative, while keeping all imperative parts application agnostic, and thus reusable across different applications. For achieving this we turn to another domain, namely, *Domain-Specific Languages (DSLs)*.

DSLs can be implemented as either external DSLs, implemented as compilers or interpreters, or internal [10, 34] or embedded [23, 21] DSLs, implemented as software libraries in a host language, in a process sometimes referred to as *DSL embedding*. For declarative host languages, embedded DSLs are defined declaratively [23]. This results in a desired separation of code assets, where all DSL-specific code (namely the DSL definition) is declarative, and the (often imperative) implementation of the host language remains DSL-agnostic, and therefore reusable across DSLs.

We can apply DSL embedding to web applications by thinking of the state of applications as programs in a languages. The equivalence between code and data is a fundamental concept in Computer Science (applied, e.g., in Von Neumann's stored-program architecture [38] and in Church's lambda calculus [4]). This equivalence is often used to describe code as data (e.g., an encoding of a program), but in some cases the opposite direction is useful as well. In our case, we note that the logic-tier of an application following the 3TA acts as an interpreter, acting upon the "program" stored in the data-tier as the state of the application. The database schema defines the language this program is written in, while the business logic gives it meaning, acting as its semantics. Since the concepts stored in the database refer to the application's problem domain, its corresponding language can be treated as a DSL.

By drawing this mapping between applications and DSLs we open the door to comparing their implementation techniques. Indeed, state-of-the-art techniques for web application development typically compare to external DSLs, while methods that map to DSL embedding are seldom used for anything other than static websites.

### 1.3 Contribution

We present *Application Embedding*—a web application development approach based on DSL embedding, in which a web application is embedded in a host application, just like a DSL is embedded in a host language. Just as in the field of DSLs the host language is a general-purpose programming language, our host application is a *general-purpose application*, fully programmable to host any embedded web application. As in DSLs, this approach results in a desirable separation of code assets. The imperative implementation of the host application is reusable, since it is general-purpose, while all the application-specific code is declarative, consisting of definitions in the language implemented by the host application.

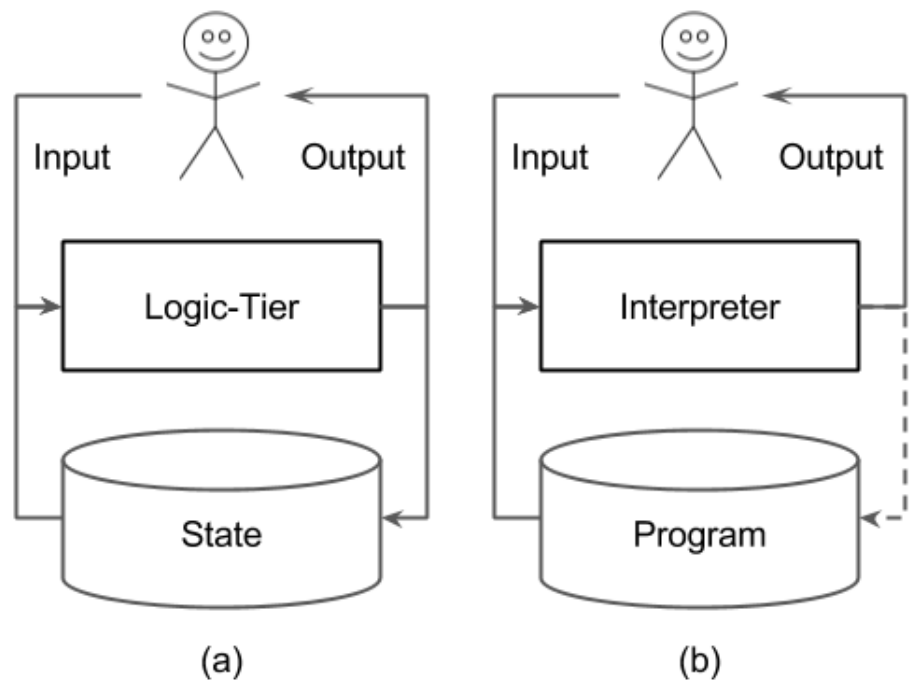

**Figure 1** A comparison between (a) the logic-tier of an application and (b) an interpreter

As validation of this approach, we present the following:

- A discussion of the mapping between web applications and DSLs, demonstrated by providing specifications for a web application using formal semantics of the corresponding DSL (section 2).
- A discussion of the consequences of applying this mapping to embedded DSLs, deriving requirements for *host applications*, needed to implement our *application embedding* approach (section 3).
- A computational model, named the *Fish-Tank Model (FTM),* that is based on *Logic Programming* (*LP*) and meets these requirements, along with an implementation of this model, name `FishTank` (section 4).
- A purely declarative definition of a microblogging web application, embedded in `FishTank` (section 5).

## 2 An Application as a Programming Language

Since the early days of Computer Science it is known that programs and their state are interchangeable. Von Neumann's stored-program architecture [38] requires programs to be encoded as data. By this architecture, a program, like its state, is a part of the state of the machine running it. Curry's lambda calculus [4] goes beyond that, making it possible to treat the state of any program as a program by itself.

While it may be confusing to think of state as a program in the general case, it is much easier to see the correspondence in applications that follow the 3TA. Figure 1 compares a logic-tier of an application (figure 1a), to an interpreter (figure 1b) of a programming language. Both logic-tiers and interpreters process input from a user to

<table>
<tr><th></th><th colspan="2">Application: App</th><th colspan="2">Language: $\mathscr{L}\langle App\rangle$</th></tr>
<tr><td rowspan="4">(a)</td><td rowspan="4">Special-purpose</td><td>(Database) Schema</td><td>Abstract Syntax</td><td rowspan="4">DSL</td></tr>
<tr><td>Business Logic</td><td>Semantics</td></tr>
<tr><td>Presentation</td><td>Concrete Syntax</td></tr>
<tr><td>State (Data)</td><td>Program</td></tr>
<tr><td rowspan="2">(b)</td><td rowspan="2">Traditional</td><td>Web App</td><td>External DSL</td><td rowspan="2">External</td></tr>
<tr><td>Logic-tier</td><td>Interpreter</td></tr>
<tr><td rowspan="2">(c)</td><td rowspan="2">Embedding</td><td>Embedded App</td><td>Embedded DSL</td><td rowspan="2">Internal</td></tr>
<tr><td>Host App</td><td>Host Language</td></tr>
</table>

**Table 1** Comparing Applications with Languages

produce output. Both do so according to their own programming: the logic tier follows the application's business logic and the interpreter follows the programming language's semantics. The programming is complemented by additional input: the application's state (stored in the data-tier) or the program being interpreted, respectively. In the case of the application, however, state is also updated by the logic-tier, while interpreters typically do not modify the program they run.

This comparison is summarized in table 1. For an application *App* we denote the corresponding language by $\mathscr{L}\langle App\rangle$ (dubbed "*App* language"). The database schema used by *App* provides the abstract syntax for $\mathscr{L}\langle App\rangle$. *App*'s presentation (user interface) gives human-readable and writable form to its state, acting as its concrete syntax. Most applications are *special-purpose*, built around one specific problem domain (e.g., social media, e-commerce). In such applications the state consists of concepts from that problem domain (e.g., tweets, statuses, products). This makes their corresponding languages domain-specific, i.e., DSLs (table 1a).

The rest of this section relates the different aspects of an application to the corresponding aspects of a DSL. To demonstrate this correspondence we describe the behavior of an example web application named TweetLog via a specification of its corresponding $\mathscr{L}\langle$TweetLog$\rangle$ DSL. TweetLog is a microblogging application, where users can publish short textual *tweets* and *follow* other users. The application then presents users with *time-lines*, containing tweets made by users they follow.[6]

### 2.1 Schema as Abstract Syntax

For an application *App*, the schema of the data stored in *App*'s database corresponds to the abstract syntax for $\mathscr{L}\langle App\rangle$, where each program $P \in \mathscr{L}\langle App\rangle$ corresponds to a single state. It is common practice to treat the state of a database at a certain point in time as a mathematical set, where each element in that set represents a single piece of information (record, document, etc). We therefore treat $P$ as a *set of statements*. Therefore, the abstract syntax of these statements defines the structure of the state.

[6] We provide only anecdotal definitions here. The complete specification can be found in appendix A.

To define the schema of TweetLog's state we define the abstract syntax of its corresponding $\mathscr{L}\langle\text{TweetLog}\rangle$ DSL:[7]

$$\begin{array}{rcl} \mathrm{S} & = & \mathit{tweeted}\,(\mathrm{User}, \mathrm{Time}, \mathrm{Tweet}) \\ & | & \mathit{follows}\,(\mathrm{User}, \mathrm{User}, \mathrm{Time}) \\ & | & \mathit{timeline}\,(\mathrm{User}, \mathrm{Time}, \mathrm{TimelineElem}) \\ & | & \ldots \\ \mathrm{User} & = & \mathit{user}\,(\mathrm{String}) \\ \mathrm{Tweet} & = & \mathit{text}\,(\mathrm{Text}) \\ & | & \mathit{image}\,(\mathrm{Binary}, \mathrm{Text}) \\ & \ldots & \end{array} \tag{1}$$

Statements in $\mathscr{L}\langle\text{TweetLog}\rangle$, denoted by the S symbol, are tweets (*tweeted* statements) and following relations (*follows* statements), which are both considered *raw* statements, since users add these statements explicitly. The abstract syntax also accounts for *derived* statements, such as time-line entires (*timeline* statements), which are calculated by the business logic. These statements may or may not be actually stored by the application, but they can all be inferred based on the state of the application, and thus we consider them part of the state.

### 2.2 Business Logic as Semantics

The business logic of an application *App* can be thought of as both the static and dynamic semantics of $\mathscr{L}\langle App\rangle$. The business logic serves two major roles in an application (which are not mutually exclusive): it *mutates* the state of the application and *acts upon it*. In its former role, the business logic actually refines the set of allowed states, beyond what the schema allows. This corresponds to the role of the static semantics in a programming language, which refines the set of allowed programs beyond what the abstract syntax allows. In its latter role, the business logic provides the meaning of the application state. This corresponds to the role of dynamic semantics, which provides meaning to programs.

Out of the two roles, we describe here only the dynamic semantics of $\mathscr{L}\langle\text{TweetLog}\rangle$. For simplicity we assume that any set of syntactically-correct statements is a legal program in $\mathscr{L}\langle\text{TweetLog}\rangle$, and thus a legal state in TweetLog. Real-life applications will guard against certain kinds of syntactically-valid state transitions to maintain data integrity. A method for providing integrity, as well as confidentiality in similar settings is provided elsewhere [28].

**Notation** To define the dynamic semantics for $\mathscr{L}\langle\text{TweetLog}\rangle$, we use a system of natural deduction rules as a way to reason about logical judgments relating programs to their input and output. In Natural Semantics [25] judgments usually correspond to typical behavior of general-purpose programming languages, such as state transition of imperative languages or the evaluation of an expression to a value in functional languages. $\mathscr{L}\langle\text{TweetLog}\rangle$, however, is neither imperative nor functional. Our way of reasoning upon it is by stating what statements S are true given program $P$, denoted $P \models \text{S}$. First and foremost, a statement S is true given program $P$ if it is *included* in $P$:

$$\frac{\text{S} \in P}{P \models \text{S}}$$

[7] The complete abstract syntax specification for $\mathscr{L}\langle\text{TweetLog}\rangle$ can be found in figure 10.

The following rules extend this set of statement with derived statements. In addition to $\models$, we use ad-hoc judgments, which we denote as a first-order logic atom, using a name and arguments.

**Semantic Rules** We begin by tokenizing each tweet to produce a processed tweet (*procTweet*):

$$\frac{\begin{array}{c} P \models \mathit{tweeted}(U,\tau,T) \\ \mathit{parse}\left(\mathit{tokens}\left(X'\right),X,\varepsilon\right) \\ \mathit{replaceText}\left(T,\mathit{plain}(X),\mathit{tokenized}\left(X'\right),T'\right) \end{array}}{P \models \mathit{procTweet}\left(U,\tau,T'\right)}\ \textsc{proc-tweet} \tag{2}$$

Here a processed tweet is based on a "raw" tweet, after extracting its text and parsing (tokenizing) it. The *parse* judgment asserts that $X$ can be tokenized into a list of tokens $X$', and *replaceText* replaces the plaintext $X$ with the list of tokens $X$' in the body of the tweet.[8]

This tokenization is then used to create a search index (*searchIndex*), indexing tweets by their tokens (simple words, user IDs, and hashtags):

$$\frac{P \models \mathit{procTweet}(U,\tau,T) \quad \mathit{replaceText}(T,\mathit{tokenized}(X),\mathit{tokenized}(X),T) \quad \xi \in X}{P \models \mathit{searchIndex}(\xi,\tau,T)}\ \textsc{search}_1 \tag{3}$$

Additionally, tweets are also indexed by the user making them:

$$\frac{P \models \mathit{procTweet}(U,\tau,T)}{P \models \mathit{searchIndex}(\mathit{userID}(U),\tau,T)}\ \textsc{search}_2 \tag{4}$$

The search index can be used for regular searches, but in particular it is used to define the time-line:

$$\frac{P \models \mathit{follows}\left(U_1,\mathit{user}(U_2),\tau'\right) \quad P \models \mathit{searchIndex}(\mathit{userID}(U_2),\tau,T)}{P \models \mathit{timeline}(U_1,\tau,\mathit{tweet}(U_2,T))}\ \textsc{time-line}_1 \tag{5}$$

By this, a user's time-line for user $U_1$ consists of all tweets related to any user $U_2$ followed by $U_1$. This includes both tweets made by $U_2$ as well as all replies to these tweets, i.e., tweets in which user $U_2$ is mentioned.

[8] The definition of *replaceText* is given in equation (7) and equation (8). The definition of *parse* is given in figure 12.

### 2.3 Presentation as Concrete Syntax

For an application *App*, the concrete syntax of $\mathscr{L}\langle App\rangle$ defines how programs (state of application *App*) is presented to the user. In programming, we are used to thinking of concrete syntax in the other direction, defining how textual source code translates to an abstract syntax tree. In contrast, here we consider concrete syntax as a bidirectional mapping between the abstract and concrete representations of a program. *Projectional Editing* (*PE*) [18, 39, 14], for example, is a technique by which concrete syntax defines how an *Abstract Syntax Tree* (*AST*) is projected to a view displayed to the user. Apart for displaying the AST, the view also facilitates editing of the AST. These edits are made visible to the user by applying the projection again, to update the view.

In DSLs that represent web applications the meaning of concrete syntax is similar to its meaning in projectional editing. The view in this case consists of HTML on a web browser, and the projection translates the abstract representation (data in the database) into its HTML representation. The view is also responsible for mutating the abstract representation by allowing the creation of new data, and editing or deleting data.

Recall that the state of the application consists of a set of statements. This means that mutation of the state includes adding new statements, deleting statements, and modifying statements. HTML provides *forms* to perform addition of new data (statements) and editing existing data. Delete buttons can be associated with data to be displayed to facilitate deletion.

### 2.4 State as Program

The mapping of the state of an application to a DSL program is somewhat counterintuitive at first. While programs are typically immutable and relatively small, the state of an application is constantly mutated by the logic-tier and is typically very large. In this section we elaborate on these differences and provide intuition as to which languages map well to applications.

**State as a set implies a declarative program** One thing to notice about $\mathscr{L}\langle App\rangle$ is that it is purely declarative for almost any *App*. Recall that program $P \in \mathscr{L}\langle App\rangle$ is an unordered set of statements. Such a set cannot represent a sequence of commands to be executed (imperative code), unless individual statements do. Statements can represent such sequences of commands in applications that allow users to build such sequences, such as programming environments, where the application stores and executes user programs. In any other case, we can refer to $\mathscr{L}\langle App\rangle$ as a declarative language.

**Mutable state implies a mutable program** Recall program $P \in \mathscr{L}\langle App\rangle$ represents the state of application *App* at a given moment. This makes $P$ transient—applicable to a single, passing moment. The facts the application's logic-tier mutates the state corresponds to the interpreter of $\mathscr{L}\langle App\rangle$ mutating the program. This mutation is depicted as a dashed line in figure 1b.

Program mutation is considered hard in imperative languages, because mutating a program as it runs requires a corresponding restructuring of its state. However, declarative programming languages do not have a notion of state, making them easier to mutate. In fact, some declarative paradigms have a long tradition of program mutation as alternative to storing state. The Prolog [8, 12] predicates assert/1 and retract/1, for example, add and remove clauses to and from the *program*. Many imperative Prolog programs that interact with users modify a declarative subset of the program using these predicates in order to store user data. Then, to answer user questions, they query that declarative subset [5, p163].[9]

**Big state implies a large program** Most implementations of traditional programming languages make assumptions regarding the size of the program. Regardless of whether the language is compiled or interpreted, a common assumption is that a program is small enough in size so it can be loaded completely into memory. In contrast, application developers typically do not make similar assumption regarding the size of the data (state) stored by the application, and such data is often stored to disk on the database, and in some cases is even distributed across many database nodes. Consequently, traditional language implementation techniques are often inadequate for implementing applications.

However, some existing languages can handle such large programs. One example is Datalog [6]—a data representation and query LP language for deductive databases [33]. It is not Turing complete, and thus is not a general programming language, but its "programs" (the contents of a deductive database) are both declarative, large, and mutable.

## 3 Application Embedding

Having established the similarity between applications and DSLs, next we examine the two well-known techniques for implementing DSLs, and correlate them to different approaches for application development.

### 3.1 Embedded Applications as Embedded DSLs

In the field of programming languages, there are two well-known techniques for implementing DSLs: *external DSLs* are implemented as compilers or interpreters, while *internal DSLs* [10, 17, 30, 34] are implemented as software libraries in a host language in a process sometimes referred to as *DSL embedding* [23, 21], such that a program in the DSL is also a program in the host language.

The two approaches have significant trade-offs [19, 27]. On the one hand, external DSLs give their developers absolute freedom in defining the syntax and semantics of the DSL, while internal DSLs are bound to restrictions (both syntactic and semantic) posed by the host language. On the other hand, external DSLs leave the burden of providing an efficient implementation, as well as editors, debuggers, and other tools on the DSL developer, while internal DSLs reuse much of the facilities created for the host language. For example, an optimizing compiler that exists for the host language will optimize (with varying levels of success) code written in DSLs it hosts.

[9] An example Prolog program that uses mutating as an alternative to manipulating state is presented in appendix D.

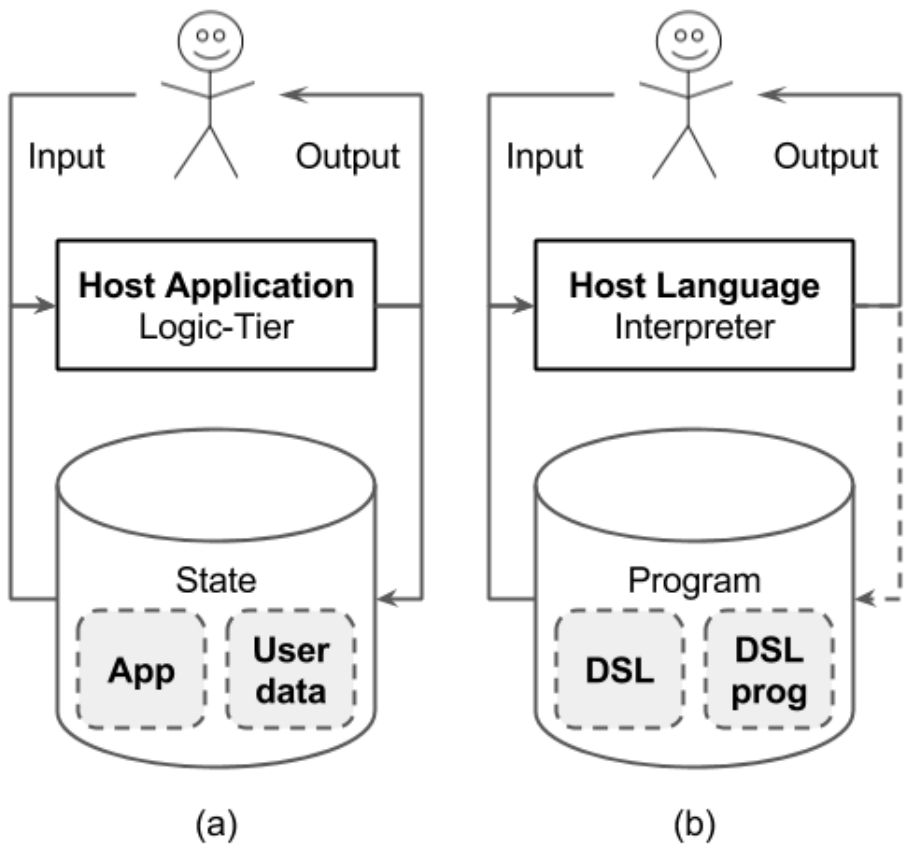

**Figure 2** A comparison between (a) a host application and (b) a host language

**Web applications as external and internal DSLs** Recall that we compare the logic-tier of a web application to an interpreter of a DSL (section 2). Consequently, applications implemented using the 3TA correspond to external DSLs, implemented as interpreters. Indeed, applications implemented in languages such as PHP [26], Java [2], Ruby [37], or JavaScript [16] are strictly external to their respective implementation language. In contrast, websites built using a *Content-Management System* (*CMS*) such as WordPress[10] or Drupal[11] can be seen as *embedded* in these applications.

We coin the term *Application Embedding* to describe the implementation of an *embedded application* as payload over a *host application*. Figure 2 augments figure 1 to describe how internal DSLs map to embedded applications such as WordPress websites. WordPress itself is a standard 3TA web application for which the logic-tier acts as an interpreter over its state. The state, however, consists of mainly two parts: definitions of guest applications (e.g., user website content), and application user-data (e.g., comments left on user websites). The two live side-by-side as the state of WordPress—the host application. This is similar to a Ruby program consisting of a DSL definition provided as a library [10] and a DSL program, both living side-by-side as Ruby code, interpreted by the Ruby interpreter which is implemented externally of Ruby (table 1c).

As expected, traditional and embedded web applications feature similar trade-offs to the ones existing for external and internal DSLs. Building a website using a CMS is usually much easier than building it using, e.g., a web framework. However, a web framework gives much more freedom in both the application's visual design (its "syntax") and its functionality ("semantics"). In both cases, CMS users are bound to what the particular CMS provides. Specifically, websites built using CMSs are typically almost entirely static, where the user building the website can only choose functionality from a predefined collection (a contact form, comments, "likes", etc).

[10] https://wordpress.org/
[11] https://www.drupal.org/

### 3.2 Host Application as a Host Language

Following CMS as a conceptual example, a host application needs to be, first and foremost, a web application. It needs to store state, which consists of both the embedded application (e.g., the content of a website in a CMS), along with the state of that application (e.g., comments and likes on pages of such a website). However, unlike CMSs, we want our host application to support arbitrary functionality. Therefore, the host application needs to be general-purpose. We define a *general-purpose application* as an application *App* for which $\mathscr{L}\langle App \rangle$ is a general-purpose programming language. We can therefore start by discussing requirements for a host language, and then move on to discussing a corresponding host application.

**A host language** Datalog is an example of a language that can represent large, mutable, declarative programs (section 2.4). Unfortunately, two problems hinder the use of Datalog as a general-purpose host language. First, it is not Turing-complete. As a result it is not suitable for expressing arbitrary functionality. Second, the well-known method for embedding DSLs in LP languages [29] is inapplicable to Datalog, since it does not support compound terms. An appropriate host language should, like Datalog, be declarative and support large and mutable programs; but, unlike Datalog, it should be Turing-complete, and support DSL embedding.

In previous work [28] we introduced the CloudLog language. Like Datalog, CloudLog is a LP data representation and query language. However, unlike Datalog, CloudLog is Turing-complete and provides a choice between Prolog-like *(top-down) clauses*, evaluated at query time, and Datalog-like *(bottom-up) rules,* evaluated at update time. Running logic at update time, a practice known as *denormalization*, is common in applications that handle large datasets and particularly in conjunction with NoSQL databases [15, 31]. In this paper we show that CloudLog is also capable of hosting internal DSLs.

**A host application** A host application *hostApp* is an implementation of a corresponding host language $\mathscr{L}\langle hostApp \rangle$. If our host language is a data representation language, then the corresponding host application is a database. Specifically, if the host language is a LP data representation language such as CloudLog, the corresponding host application is a deductive database. In section 4 we present a concrete host application, FishTank, such that $\mathscr{L}\langle \text{FishTank} \rangle = \text{CloudLog}$. In addition to being a database, a host application must also be able to store and serve static content (HTML, CSS, images, and JavaScript files) which it receives from developers, and provide access to the database through HTTP calls.

## 4 FishTank: A Concrete Host Application

To validate our approach we introduce FishTank, a host application we have implemented as proof-of-concept, and explain its underlying LP computational model—the *Fish-Tank Model (FTM)*. The novelty of FTM is not in the result of the computation, which is a derivation of theorems from axioms through *Modus Ponens (MP)*, but rather its unique imperative behavior, namely an explicit choice by the users of this model between update-time and query-time evaluation for each rule.

For the sake of clarity, we describe this model in three steps, starting with a simplified model that only supports update-time evaluation (section 4.1), continuing to the complete theoretical model also supporting query-time evaluation as well as application of arbitrary logic at update time (section 4.2), and finally, posing constraints on the theoretical model to make it adequate for an efficient implementation (section 4.3). After that we discuss the design and implementation of FishTank, our proof-of-concept implementation of this model (section 4.4).

### 4.1 A Simplified FTM

FTM is a computational model based on LP.[12] It is based on the notion of a container (the "tank") that, at any given point in time, contains a set of logic axioms (the "fish"). These axioms can be *facts* (male fish)—first-order-logic atoms, *rules* (female fish) of the form $F \rightarrow A$ where $F$ is an atom (corresponding to a fact) and $A$ is an axiom of any kind, and *clauses* (infertile fish), which are discussed in section 4.2.[13]

When a fact is added to the fish-tank, all matching rules are located and applied to that fact through MP. The resulting axiom from each match is recursively added to the fish-tank. Similarly, when a rule is added, it is applied to all matching facts in the fish-tank, and all resulting axioms are added. When a fact or a rule are removed from the tank all descendant axioms are removed as well.

**Properties** This simple model has a few important properties. First, it is *declarative*. The symmetry between facts and rules, in conjunction with the symmetry between insertion and removal, guarantees that when this system comes into a steady state, the contents of the container is a closure (over MP) of the original set of axioms that were inserted but not removed, regardless of the order in which these operations were applied. Second, under some simplifying assumptions discussed in section 4.3, this model can be implemented to provide efficient storage and retrieval for large datasets. Retrieval starts with a query term (an atom), looking for either facts that are potentially unifiable with it, or rules that can be applied to such facts. Retrieval is the key for both querying the state of the container and for finding possible matches for inserted rules or facts. Third, this model is scale-friendly. Since it does not provide global atomicity, there is no single point of failure or no inherent global bottleneck limiting scalability.

[12] We call this model the *Fish-Tank Model* due to the spontaneous nature in which facts and rules interact to create new axioms, like male and female fish spontaneously finding each other to produce offsprings in a fish-tank.

[13] Here we give an intuitive high level description of the computation model. The precise and complete definition of the FTM is provided in appendix B.

**Limitations** The question remains, is this model powerful enough to represent the business logic of any application? The simplified FTM presented so far is somewhat limited. Consider for example a simple search engine built on top of the FTM. The input to this search engine is a set of facts representing documents that contain text, and the output is a result for a query that is given one or more keywords. The simplified FTM is missing two key capabilities to allow an efficient implementation of such an engine. The first is the ability to perform arbitrary logic (in this case, extraction of keywords from the text) when applying rules to facts. While such logic can be performed in our model, it will require the creation of at least one new axiom per each step of the parsing process, i.e., for each character in the text. Even if keywords are successfully extracted, multi-keyword searches (i.e., queries for documents containing all keywords within a given set) are inefficient under the simplified FTM. As we do not have access to the query itself when we prepare the results, it is not enough to index documents by each keyword they contain. We need to index each document by each possible combination of keywords. Even if we limit the number of keywords in a search to $k$, we still need to have $\binom{n}{k}$ index entries *per document* (where $n$ is the number of different words in the document). If we could apply logic at query-time, we could have held an index mapping single keywords to documents ($k$ entries per document), and when searching for a combination of $k$ keywords we would search for only one keyword, and then filter out results not containing the remaining $k-1$ keywords.

### 4.2 The Complete FTM

The complete FTM addresses the need for query-time logic by introducing *dynamic clauses, predicates,* and *goals*, and addresses the need for arbitrary logic at propagation-time by introducing *guards*.

**Dynamic clauses, predicates, and goals** Facts and rules exist in the FTM to create dynamic clauses, which are consulted by queries. Dynamic clauses are axioms of the form $H \vdash B$, where $H$ is an atomic dynamic goal, and $B$ is a dynamic goal of any kind. Clauses, like rules, convey logical deduction. However, while rules are applied at update time, clauses are applied at query time. Dynamic clauses define dynamic predicates—predicates representing relations that change over time. This is in contrast to static predicates, which once defined, do not change. A dynamic predicate is expected to consist of a large number of clauses, each providing a separate set of results. In many cases, many of these clauses will be trivial, i.e., of the form $H \leftarrow \top$, where $\top$ is the trivial goal. These are the cases where the propagation of rules and facts applied all the necessary logic. However, in cases where logic needs to be applied at query time (e.g., in multiple-keyword searches), a non-trivial body ($B$) defines the steps necessary to find the results.

**Guards** In cases where non-trivial logic needs to be applied during the propagation of axioms, a *guard*, can be provided as a part of the rule. A rule is defined as $F\{G\} \to A$, where $G$ is a goal representing the guard (a rule of the form $F \to A$ is read as $F\{\top\} \to A$, where the guard is the trivial goal $\top$). During the application of a rule to a fact, the fact's guard is evaluated, and a resulting axiom is produced *for every result of the guard*. If the guard fails, no axioms will be produced even though the fact matched the left-hand side of the rule. If the guard succeeded multiple times, multiple axioms will be produced. In section 5 we use guards to index tweets in TweetLog. The guard examines tweets and provides a result for each token in each tweet. As result, the rule produces a search result for each such token.

Guards are required to be *static* goals, i.e., goals for which the set of results does not change over time, to ensure consistent behavior between additions and removals. When a fact is removed, all corresponding rules are applied to it to find out which other axioms also need to be removed. For each such application, the guard is consulted. If the guard behaved differently than the way it originally behaved when the fact (or the rule) were introduced, the removal would not have been complete.

### 4.3 FTM Made Practical

The complete FTM is a theoretical model. To make it practical and efficient, some simplifying assumptions on usage need to be made.

**Concreteness and indexing** We distinguish between two kinds of axioms: *concrete* and *generic*. *Concrete facts* are facts of the form $f(t_1, \ldots, t_n)$, where $t_1$ is a *ground term* (a term that does not contain logic variables). A *concrete rule* is a rule of the form $F\{G\} \to A$ where $F$ is a concrete fact. A *concrete clause* is a clause of the form $p(t_1 \ldots t_n) \vdash B$, where $t_1$ is a ground term. *Generic axioms* are axioms that are not concrete.

To promote efficient implementation we assume that there is a small number of generic axioms at any given time in the fish-tank, and that they are seldom added or removed. Intuitively, such axioms should be considered a part of the embedded application's definition, and not a part of its state. The state should consist exclusively of concrete axioms. We also assume that the number of facts, rules, or clauses in the fish-tank, with a certain $t_1$ value, is bounded at any given time.

Concrete axioms are indexed by the value of $t_1$—their *subject*. When a concrete fact $f(t'_1, \ldots, t'_n)$ is added to the fish-tank, matching rules are found by using $t'_1$ as key. When a rule is added, matching facts are found the same way. Similarly, concrete queries will only consider concrete clauses they will find using the goal's first argument.

Generic axioms are harder to manage. Having variables in their first argument (or having a variable *as* first argument) there is no easy way to retrieve a bounded collection of matching axioms. As result, any update (addition or removal) of generic facts or rules requires processing all rules or facts (respectively) that share the same fact-name ($f$). While this process may take some time, it does not happen often. It occurs on software updates and resembles data migration that traditionally occurs in such events.

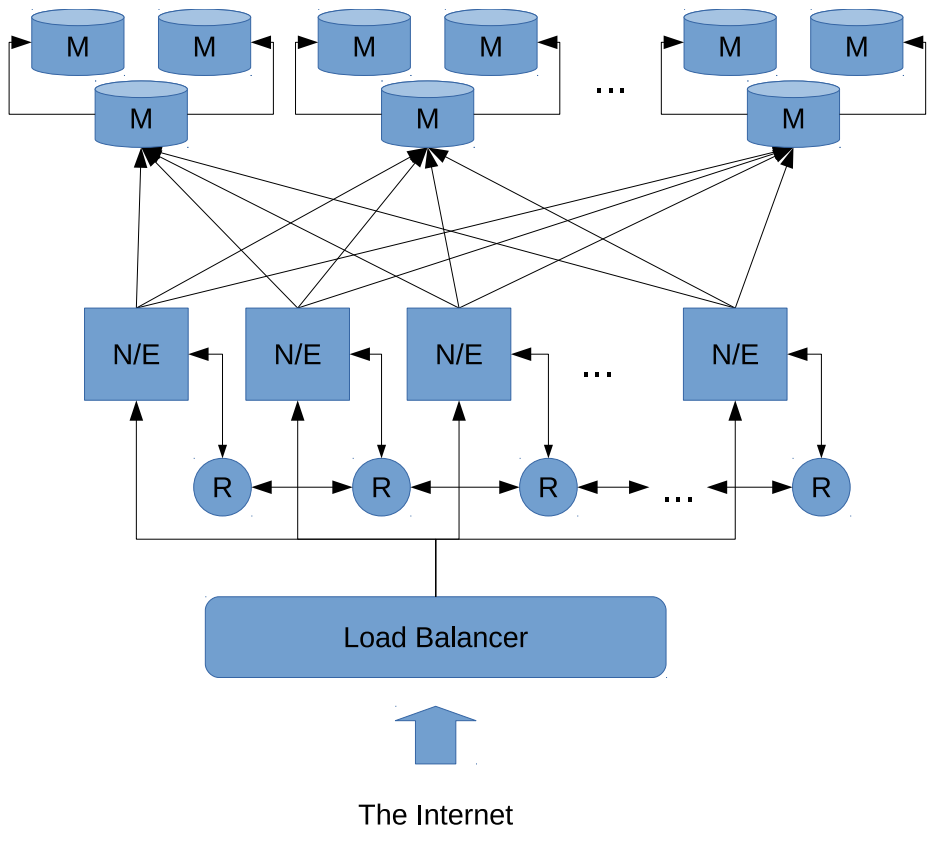

**Figure 3** FishTank Architecture. R, N/E, and M respectively denote RabbitMQ, NodeJS/Express, and MongoDB.

### 4.4 Design and Implementation

Our proof-of-concept implementation of the FTM is called FishTank.[14] FishTank's architecture is based on MEAN [11]: MongoDB [3], Express,[15] AngularJS, and NodeJS.[16] This is a relatively new architecture for web applications that is designed for scale (the use of the NoSQL database MongoDB), and a single implementation language: JavaScript. We use an architecture designed for web applications because FishTank is a web application.

Figure 3 shows the structure of a typical FishTank cluster. It consists of a MongoDB cluster (M-nodes), a RabbitMQ cluster (R-nodes) and an array of NodeJS/Express nodes (N/E nodes) connected to a load-balancer.

The MongoDB cluster stores the axioms along with their multiplicity.[17] All facts with a certain first argument value, and all rules matching such facts are stored in the same MongoDB document. MongoDB guarantees atomic updates within a single document. This is important to avoid the chance that a fact and a rule, added to the database nearly at the same time, miss each other during propagation.

Axioms resulting from applying rules to facts are not added directly, but instead are placed in a common work queue, maintained by the RabbitMQ cluster.[18] We do this to ensure low update latency. By placing these axioms in a highly-available queue we guarantee that these axioms will eventually be added to the database without having the user wait until this happens.

**Scalability and availability** FishTank's design follows the 3TA, with a separation, on the server-side, between the stateless N/E nodes, and the stateful (but application-agnostic) M-nodes and R-nodes. This separation allows us to use a large number of N/E nodes to address high load. Being stateless, they become redundant, so that if one of them crashes the load balancer can send subsequent requests to the other nodes and the service as a whole will remain available. MongoDB and RabbitMQ are both designed for scalability and availability, so the entire system will not fail due to a failure of a single M-node or R-node, or even multiple nodes (in most cases).

[14] https://github.com/brosenan/fishtank
[15] http://expressjs.com/
[16] https://nodejs.org/
[17] Corresponds to $T$ in appendix B.4.
[18] Corresponds to $Q$ in appendix B.4.

**Implementation** Being based on the MEAN architecture, JavaScript plays an important role in the implementation of FishTank. FishTank is implemented as a server (over NodeJS and Express), and a client library, extending AngularJS. The server-side code is divided to JavaScript code, providing capabilities such as responding to HTTP requests and using MongoDB and RabbitMQ, and Cedalion [27] code, implementing the logic itself. We chose Cedalion due to its expressiveness and due to the facts it is a LP language, making it a natural choice for implementing a deductive database.

To give a sense of size, the server-side implementation consists of ~1000 Cedalion *statements* (a typical Cedalion statement represents a few lines of code, but Cedalion uses projectional editing, so the classic measure of lines-of-code does not apply), and ~3000 lines of JavaScript code. Both include unit tests, which on the JavaScript side accounts for about half the code. The client-side library consists of ~650 lines of code, about half of which are unit tests. However, this library is currently very partial, implementing just enough functionality to support the implementation of TweetLog.

## 5 TweetLog: *An Embedded Application*

We evaluate our approach by demonstrating that: (1) the FTM is powerful enough to allow non-trivial applications to be developed over it, and (2) the proposed separation of code assets successfully segregates all imperative code in the application-agnostic implementation of FishTank, so that all application-specific assets (on the server, and with small unavoidable exceptions, on the client) are declarative.

We demonstrate both by presenting an implementation of TweetLog[19] over FishTank. We show that this implementation closely relates to our definition of the corresponding $\mathscr{L}\langle$TweetLog$\rangle$.

Figure 4 shows a screenshot of TweetLog running on a web browser. The navigation bar at the top of the page allows users to log in (enter their user name) and to search for content. The page displayed in figure 4 is the timeline of user Foo, which is the page a user enters after logging in. This page shows the tweets the user registered to, and provides boxes for tweeting and for following users. On the left, the sidebar menu contains links to the timeline (this page), a page showing the tweets of the logged-in user, his or her followers and followees (followers and followees are not yet implemented), as well as an *about* page describing the application.

[19] https://github.com/brosenan/tweetlog

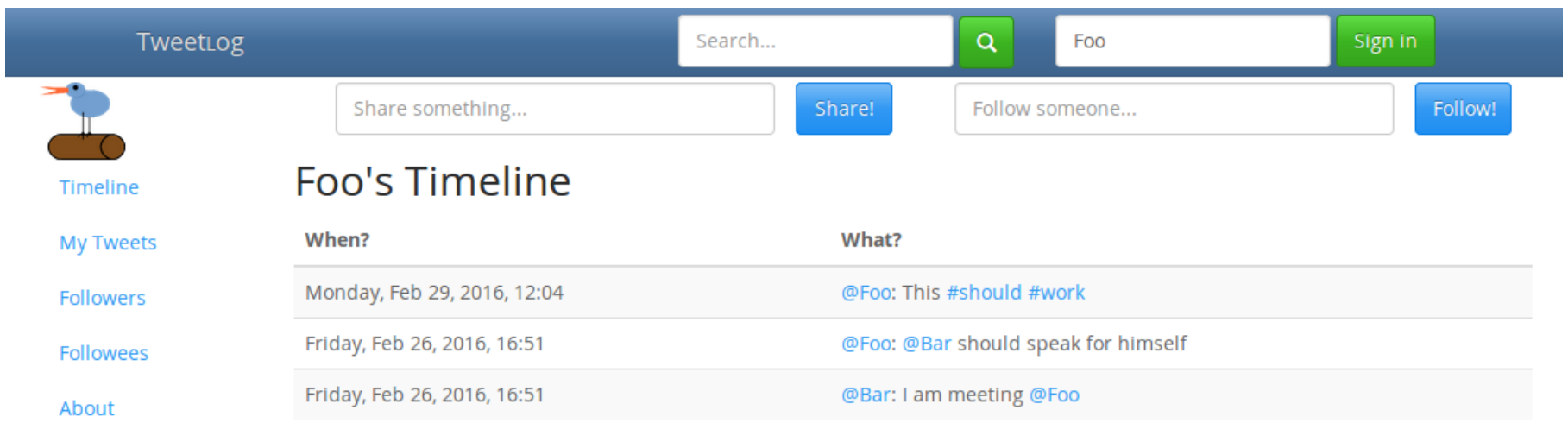

**Figure 4** Screenshot of the TweetLog application

```
1. declare user :: type
2. declare @ Name :: user
      where  Name :: string
      display as h  "@"   《 Name :: string 》
```

**Figure 5** User Abstract Syntax

Tweets are tokenized into words, hashtags, and user IDs. User IDs and hashtags become links to pages showing tweets related to a user or containing a hashtag, respectively.

### 5.1 Abstract Syntax

We use Cedalion concept declarations [27] to define concepts in the abstract syntax of $\mathscr{L}\langle$TweetLog$\rangle$. For example, figure 5 shows a piece of $\mathscr{L}\langle$TweetLog$\rangle$'s abstract syntax. Statement 1 in figure 5 declares the type User, and statement 2 in figure 5 is equivalent to User $=$ *user* (String) from equation (1). This declaration has two parts: a mandatory type signature ("where Name::string") and an optional projection definition ("display as…"). That second part gives the *user* concept its concrete syntax within the *context of* $\mathscr{L}\langle$TweetLog$\rangle$'s *definition*, and allows it to display as *@Name* rather than *user* (*Name*). This projection definition helps making $\mathscr{L}\langle$TweetLog$\rangle$'s definition more concise and intuitive. However, in our current implementation we chose not to impose this concrete syntax on $\mathscr{L}\langle$TweetLog$\rangle$ users, and thus this concrete syntax is separated from the concrete syntax of $\mathscr{L}\langle$TweetLog$\rangle$, which is described next.[20]

### 5.2 Concrete Syntax

The concrete syntax of $\mathscr{L}\langle$TweetLog$\rangle$ is defined using templates embedded in the application's main HTML file (index.html). Figure 6 shows the template defining the concrete syntax of *user* (String). It comes as a **<script>** element, of type `ng–template`, which is picked up by AngularJS. We add the `clg–concept` attribute to associate this template with the *user* concept from the abstract syntax. The user name is bound to User, that is used inside the template.

[20] The complete abstract syntax can be found in figure 10.

```
<script type=”ng—template” clg—concept=”t:user(User)” id=”user”>
  <span><a href=”{{url(’userTweets’,␣[User])}}”>@{{User}}</a></span>
</script>
```

**Figure 6** Concrete syntax for *user* (String)

```
<script type=”ng—template” clg—concept=”t:tweet(User,␣Tweet)” id=”tweet”>
    <span clg—render—term=”User”></span>: <span clg—render—term=”Tweet”></span>
</script>
```

**Figure 7** Concrete syntax for *tweet* (User, Tweet)

```
<form>
  <clg—fact pattern=”t:tweeted(t:user(User),␣Time,␣t:text(t:rawText(Tweet)))”
            name=”tweet”
            assign=”User:␣$session.user,␣Tweet:␣text,␣Time:␣now()”>
  </clg—fact>
  <input type=”text” ng—model=”text”>
  <input type=”submit” ng—click=”add_tweet();␣text␣=␣””” value=”Share!”>
</form>
```

**Figure 8** Form providing the *tweeted* (User, Time, Tweet) fact concrete syntax

The template’s body is a **<span>** element, containing a link (**<a>**) linking to a page containing the tweets of the given user. The link is generated within the href attribute, using the {{...}} syntax. This syntax is replaced by AngularJS with the value to which the expression inside the braces evaluates to. In this case this is the url() function, which builds a URL using the given location and parameters. The text displayed in the link is a “@” followed by the name of the user.

Figure 7 shows the concrete syntax of a timeline entry corresponding to a tweet, is another example for such template. This time, the arguments (a user ID and a tweet) are compound terms, that by themselves need to be expanded using other templates. The clg—render—term directive in the empty **<span>** element tells the FishTank client library to render the term held by this attribute (either the user ID or the tweet) in the enclosing **<span>** element by applying further templates to either the user ID or the tweet.

The concrete syntax of an $\mathscr{L}\langle App\rangle$ must also provide users ways to mutate the state of an application (section 2.3). Currently, our client-side library only supports addition of new facts. These are performed using the **<clg—fact>** directive. Figure 8 shows a form in which TweetLog users can add textual tweets. The **<clg—fact>** directive takes a pattern (a CloudLog term containing the pattern of facts to be created, where logic variables represent the data that changes), a name, that is later used to submit the new fact, and a variable assignment, mapping values from the AngularJS scope to the logic variables in the pattern. In this case, the pattern represents a textual tweet. The first **<input>** element is a text box in which the tweet text is placed. It uses an AngularJS ng—model directive to associate the value of the field to the scope variable

```
1. axiom U tweeted T at τ { replaceText ( T , raw Text , tokenized ( X ) , T ' ) ,
                            tokens ( X , tweetCtx ) parses Text               } → procTweet ( U , τ , T ' )
2. axiom procTweet ( User , Time , Tweet ) { replaceText ( Tweet , tokenized ( Tokens ) , _ , _ ) ,
                                             Token :: < tweetCtx > ∈ Tokens                      } ↴
           → searchIndex ( Token , Time , User , Tweet )
3. axiom procTweet ( @ A , Time , Tweet ) → searchIndex ( @ A , Time , @ A , Tweet )
4. axiom U₁ follows @ U₂ since _ →                      ↴
           searchIndex ( @ U₂ , Time , U₃ , Tweet ) →   ↴
             timeline ( U₁ , Time , tweet ( U₃ , Tweet ) ) :- ⊤
```

**Figure 9** Timeline semantics

"text." The second **<input>** element is a submit button, which upon clicking calls the add_tweet() method, a method added to the scope by the **<clg–fact>** directive. It also clears the text.

Event handlers such as this ones are the only pieces of imperative code that exist in TweetLog's definition. Our design leaves client-side event handlers imperative since imperative programming provides the most expressive way to convey such actions. However, these pieces of code are very limited in their ability to affect the state of the application. The function add_tweet(), for example, takes no arguments, so the only thing it can convey is the desire to create a new tweet based on variables stored on the client. Other than that, event handlers are free to change the state of the client, as in clearing the text.

### 5.3 Semantics

Figure 9 shows the definition at the heart of the TweetLog application. Statement 1 in figure 9 represents the PROC-TWEET rule (equation (2)) from section 2.2. The *axiom* keyword tells us the rest of this statement is a CloudLog axiom (a rule, in this case). A rule has three parts: (1) a fact to match (in this case, a tweet); (2) a guard, which is a Cedalion goal to be satisfied for each such fact (in this case, a conjunction of a goal unpacking and repacking the text from the tweet, and a goal tokenizing the text); and (3) an axiom to be created for each result emitted by the guard (in this case, a *procTweet* fact).

Statement 2 in figure 9 indexes processed tweets by their content (the $\text{SEARCH}_1$ rule, equation (3)), and statement 3 in figure 9 indexes processed tweets by the user who made them (the $\text{SEARCH}_2$ rule, equation (4)). Note that both the tokenization and indexing of all tweets are done using rules, that operate upon insertion of tweets.

Statement 4 in figure 9 defines the timeline predicate. This is a rule that emits a rule that emits a clause (with :- replacing $\vdash$ in the syntax). The clause is trivial, with $\top$ as its body. Being dependent on two kinds of facts, this rule is similar to an SQL *join*. As in its specification (the $\text{TIME-LINE}_1$ rule, equation (5)), it provides an entry in the timeline for each tweet associated with each followed user. By choosing between a rule and a (non-trivial) clause we get a choice of temporal order. The order we choose here is to start from a following relationship (to get the followee), go through all associated tweets, and then place them in the timeline. Like indexing, this too happens as tweets and following relationships are introduced, and the only thing happening when a user searches for his or her timeline is fetching all timeline clauses associated with the user. Note that the first argument in the timeline predicate is the user, so FishTank guarantees fast ($O(1)$ I/O operations) access to it.[21]

## 6 Discussion

We have made certain simplifying assumptions concerning the manner in which FishTank is to be used, assumptions such as the number of generic axioms and the rate in which they are being updated (section 4.3). These assumptions have direct consequences in terms of applicability, correctness, and performance.

### 6.1 Applicability

Many programming platforms, such as databases and programming languages, make simplifying assumptions on usage. Prolog, for example, indexes the clauses of predicates based on their first argument [20]. In doing so the language implementation assumes predicates with a large number of clauses are defined in a way such that the the first argument differs between different clauses, so that after applying the index, only a small number of clauses (typically one or zero) is left to be considered. This assumption is translated to a guideline for developers, to choose the first argument wisely.

MongoDB assumes documents do not exceed a certain size. This restriction derives our restriction regarding the number of concrete facts and rules with a certain $t_1$ value. In MongoDB itself, this assumption becomes a guideline (one of many) for data modeling.[22]

Similar to Prolog and MongoDB, we treat the assumptions made by FishTank to be guidelines for users to structure their facts and rules in a certain way. The first argument of facts should convey their subject—an entity they describe. All facts and rules that are considered user data should relate to a subject. The cap on the number of facts and rules with a certain $t_1$ value maps to a cap on the number of facts describing that subject. For example, in TweetLog we use the user ID as the subject for timeline entries. If the number of expected timeline entries per user exceeds what is allowed in FishTank, a different subject needs to be chosen. One possibility is the pair $(user, day)$, where $day$ is the event's timestamp integer-divided by the number of seconds in a day. This creates subjects that are both small enough to comply with the restriction, but at the same time relevant in the sense that we will usually query all entries with a certain subject together.

The only kinds of facts and rules that cannot be associated with a specific entity are those that speak of data in general—the generic axioms that make up the definition of the application. The definition of an application is typically significantly smaller than its data, and changes only on software updates, which are usually significantly less rapid than user data updates.

[21] To complete the semantics, the tokenization rules (figure 11) are implemented using a pre-existing DSL (figure 14).

[22] https://docs.mongodb.com/v3.2/core/data-modeling-introduction/

### 6.2 Correctness

The correctness requirement for FishTank seems obvious: queries should be evaluated by taking into account all facts and rules added to the database (and not removed from it) until the time of the query, and should be evaluated following CloudLog's semantics. However, since we are talking about a large, distributed system, the notion of time is not exact. In a distributed system, every computer node has its own time and as a result a fact that is added to the database on one node is not known the same exact moment to all nodes. Synchronizing all nodes by means of, e.g., ACID (Atomic, Consistent, Isolated, and Durable) transactions can help, but will inhibit the scalablility of the system. We therefore require *eventual delivery*, that is, if we stop updating the database at one point, the database will come to rest in a situation where all facts and rules are taken into account by all queries.

Although this seems simple enough, there is one special case to watch for. If a fact and a matching rule are added to the database at around the same time, the new rule, which is supposed to be applied to all existing facts, might not yet be aware of the new fact being added. Likewise, the new fact may not be aware of the new rule. In such case, the fact and rule will miss each other and queries will not see their mutual product.

To overcome this problem our design uses a strongly consistent database (MongoDB). Strongly consistent databases do not guarantee global sequentiality, but do provide local sequentiality. In MongoDB's case, all operations performed on a single document are guaranteed to be sequential. We leverage this property to achieve correctness by storing all concrete facts and rules of the same subject (which are thus candidates for being matches of each other) in the same document. By doing this we make sure that when adding a fact we are aware of all matching rules and vice versa.[23]

### 6.3 Performance

table 2 summarizes the performance we require of our implementation. All user operations (adding or removing facts and queries for pre-calculated data) are completed after performing a constant-bound number of I/O operations, taking time linear by either the number of the results (in a query) or the number of rules matching the fact being added or removed (in an update). We allow an update operation to continue propagating asynchronously after notifying the user of its success, as long as we can guarantee delivery. That is, as long as we verified that the update does not conflict

[23] Strong consistency also solves the correctness problem for generic facts and rules, but the mechanism to support that is more complicated and beyond the scope of this paper.

| Operation | Synchronous | | Asynchronous | |
|---|---|---|---|---|
| | I/O Ops | Time | I/O Ops | Time |
| Add/Remove concrete fact/rule | $O(1)$ | $O(n)$ | $O(n)$ | $O(n)$ |
| Query de-normalized goal | $O(1)$ | $O(n)$ | N/A | N/A |
| Add/Remove generic fact/rule | $O(N)$ | $O(N)$ | $O(Nn)$ | $O(Nn)$ |

**Table 2** Performance requirements. $n$ is the number of axioms involved in a query or update. $N$ is the number of axioms in the database.

with any previous updates, and that the necessary actions were logged to a commit log.

We achieve that by using a reliable work queue (RabbitMQ). Adding or removing a concrete fact or rule is done by accessing a single MongoDB document (using $O(1)$ I/O operations), querying all matching rules or facts ($n$). We then apply them to one another ($O(n)$ time) to receive a set of facts and rules to be added or removed. Instead of adding or removing these directly we defer these operations to the work queue ($O(1)$ I/O operations, $O(n)$ time) and once acknowledged by the queue we acknowledge making the change.

Adding or removing generic axioms requires walking through all existing concrete facts and rules, looking for matches. We allow them to take $O(N)$ time, with $N$ representing the overall number of axioms in the database. It is acceptable for these operations to take a long time as they are only done when the $\mathscr{L}\langle App\rangle$'s definition changes, i.e., on application software upgrades. This is similar to data migration operations that are performed when upgrading the software of a traditional web applications.

MongoDB uses locks to synchronize operations done in the scope of a single document. We rely on this synchronization for correctness. However, under some conditions, these locks can result in congestion, which will increase query and update time beyond the desired asymptotic requirements listed in table 2. For example, if $m$ clients are trying to access (query or update) the same subject (a certain value of $t_1$ mapped to a single MongoDB document) at the same time, the locking itself will take $O(mn)$ time on average (waiting for $O(m)$ operations taking $O(n)$ time each to complete), exceeding the $O(n)$ requirement. However, we make the assumption that only a constant-bound number of operations is done at one time on a single subject. As further discussed in section 6.1, this assumption becomes a guideline for developers. For example, using a user ID as a part of the subject key (as in our example from section 6.1 of using $(user, day)$) can help ensure constant-bound parallel accesses, given that a single user does not make a large number of updates and queries in parallel.

## 7 Related Work

Most relevant related work include other approaches that can be considered *application embedding*, and *multi-tier* or *tierless* approaches for deploying code on multiple tiers from a single, coherent code artifact.

### 7.1 Application Embedding

Content Management Systems (CMSs, discussed in section 3.1) are host applications that allow users to build websites, which are stored as user data and served from within the CMS. Websites are typically very easy to build on top of CMSs, as the user does not need to worry about most technical aspects. However, such websites are bound to restriction posed by the CMS, which apply to both the website's graphical design and its functionality. Functionality is typically very limited in these systems, as they do not provide any programmability to users.

Apache CouchDB [1] is a NoSQL database that acts as a host application for embedded *CouchApps*.[24] It is a document store, storing JSON objects (documents). Its distinctive feature, however, in relation to other NoSQL document stores such as MongoDB, is a special kind of documents, known as *design documents*, that can be stored as a part of the state of a database. These documents contain JavaScript code that augments the database behavior. Most prominently, a design document can define *views*. Each view is defined using a JavaScript *map* function and optionally a *reduce* function. Unlike map/reduce capabilities that exist as a query feature in other NoSQL databases (such as MongoDB), in CouchDB a view is a part of the state of the database. This has interesting consequences. It allows map and reduce operations to be performed incrementally on a materialized view. This means that the view data is stored as a B+ tree, one tree per view. When a new document is added to the database the map function in each view is invoked on it to emit zero or more key/value pairs into the tree. When a document is deleted from the database the map functions are invoked once more to find view elements to be removed. The reduce function calculates aggregated values for each affected node in the tree. The result is automated denormalization of the database. To our knowledge, with the exception of FishTank, CouchDB is the only other database that has such support.

CouchDB's map functions are similar to rules in CloudLog and the FTM. They take a subset of the facts in the database (JSON documents) and map them into other facts (key/value pairs in the view tree). However, this cannot be used for *denormalized joins* such as the one in statement 4 in figure 9. There are two inhibitors: (1) derived facts (key/value pairs emitted by a map function) go into a dedicated view tree, and not the general database, and therefore do not trigger further map function invocations, and (2) rules cannot emit other rules. As a consequence, each key/value pair in a view tree is based on exactly one raw document. In this, the FTM and FishTank represent a more powerful approach.

### 7.2 Multi-Tier Programming

HOP [36] is a programming language that extends JavaScript with multi-tier features. A HOP program consists of *services*, which run on the server. Service code can contain client-side calls, with special syntax distinguishing between the two. Such code can appear in handlers for HTML events (HTML code can be embedded in HOP service code). Similarly, server-side code can be embedded in client-side code using different syntax. In both cases, HOP hides the necessary communication from the developer.

[24] http://docs.couchdb.org/en/1.6.1/couchapp/

Stips.js [32] is a tier-splitting process that splits a single JavaScript program into two separate programs, one running on the server and the other running on the client. The separation between client and server is made explicit by using special comments (@client and @server), but client code can call functions defined in the server code.

The goal behind HOP and Stips.js is to alleviate the dichotomy between the presentation-tier and the logic-tier. WebDSL [22], Ur/Web [7], and Links [9] go beyond that, and add the logic-tier to the mix, by adding support for schema definition and queries as a part of the language. An Ur/Web program, for example, will run on the logic tier, but will send JavaScript code to the presentation-tier, and SQL code to the data-tier. Its type system provides strong guarantees for the consistency between these artifact, making sure that, e.g., type validation in the presentation-tier is consistent with the type the same variable is given in the database schema.

While tierless programming aims to alleviate the dichotomy between the different tiers (e.g., between the logic-tier and the data-tier), application embedding aims to alleviate the dichotomy between the tenants of these tiers, namely the business logic and the data, by treating the business logic *as* data.

Multi-tier programming hides the immediate imperative aspects that are a direct consequence of the 3TA, namely the communication between the tiers. HOP and Stips.js allow client-side code to call server-side code (and in HOP's case, vice versa) in a way that resembles a regular function call. WebDSL, Ur/Web, and Links also provide a seamless integration to a database, in which the programmer is not concerned about creating a database connection or making SQL calls, but instead writes the application in a way that (in the case of Ur/Web and Links) weaves SQL queries in other code (e.g., in HTML fragments), or (in the case of WebDSL) associates data entities to output formatting.

However, multi-tier programming platforms do not inherently result in declarative programming. The manipulation of state remains imperative with these approaches. WebDSL includes a Java-like imperative programming language defined purely for the purpose of manipulating state. It treats entities as objects (hiding the fact they are stored on the logic-tier), but not hiding its imperative nature. Ur/Web and Links both depend on the declarative nature of SQL and HTML to allow applications to be defined in a mostly declarative manner, as long as their data model maps cleanly to a relational data model.

For example, a micro-blogging application similar to TweetLog could be defined in Ur/Web and Links in a declarative way, similar to our own implementation, as long as (1) it supported only one kind of tweets (e.g., textual tweets only), and (2) it did not need any kind of denormalization. In such settings, the timeline would have been defined using HTML wrapped in a declarative SQL SELECT query, which conveys the business logic in this case.

However, if we add a requirement to support, e.g., image tweets (as we do in TweetLog), the database schema and SQL query become significantly more complex. This is due to a problem known as the *impedance mismatch* [35]—a mismatch between what the domain object model requires, and what the relational model provides. The domain object model in this case includes polymorphism, in which textual and image tweets share some functionality and differ in some particular aspects. However, the relational model does not support polymorphism out of the box. By supporting compound terms, CloudLog does support polymorphism.

The introduction of denormalization, which is often necessary to keep query time independent of the overall size of the data, is another case where mult-tier programming does not provide an advantage. Without support for denormalization, applications written in Ur/Web, Links, and WebDSL will all need to maintain the redundant information explicitly, at update time, using imperative code. FishTank, on the other hand, performs this kind of work automatically, based on declarative rules.

## 8 Conclusion

In this paper we describe *application embedding*, a novel approach to the development of web applications. With our approach the complete application is divided into two separate applications: a host application and an embedded application. The former is implemented using general-purpose, often imperative programming languages, while the latter is written in a declarative programming language. The host application is general-purpose, making it suitable for, and thus reusable across multiple applications.

Our approach is rooted in the field of language engineering. It is based on a two-step thought process, in which the first draws a mapping between web applications and programming languages, and the second uses this mapping to find the novel equivalent of DSL embedding: application embedding.

As validation we provide an implementation of an application, in two parts. A reusable host application named FishTank, and an example embedded application named TweetLog. TweetLog's code is almost entirely declarative, consisting of a definition of the business logic in the form of logic rules, and the presentation given as HTML with directives defining data bindings. The only imperative parts are expressions associated with user events, as discussed in section 5.2. FishTank's code is much more complicated, but it is general-purpose by design, so that it can be used to implement potentially many different applications. Establishing this by embedding several other applications over it, however, is a topic left for future work.

One important aspect not addressed by our proof-of-concept implementation of FishTank is access control. We expect a real-life host application to address access control by authenticating users using their identity to derive their rights. In previous work [28] we describe how CloudLog can be extended to address access control, an important part of the business logic of applications.

Modern web applications are complicated. Developing a feature-rich, stable, usable application requires versatile skills. Today, web application developers are required to do everything. With lack of separation of concerns, the implementation of the business logic of a specific application is often tangled with, e.g., scalability, availability, and consistency considerations, which are often not easy to master. The separation of code assets presented in this paper decouples these concerns from any particular application specific concerns, thus allowing collaboration between teams and even companies (e.g., through open-source development) to build better host applications for everyone to benefit.

**Acknowledgements** This research was supported in part by the *Israel Science Foundation (ISF)* under grant No. 1440/14.

$$
\begin{array}{rcl}
\mathrm{S} & = & \mathit{tweeted}\,(\mathrm{User}, \mathrm{Time}, \mathrm{Tweet}) \\
 & | & \mathit{follows}\,(\mathrm{User}, \mathrm{User}, \mathrm{Time}) \\
 & | & \mathit{timeline}\,(\mathrm{User}, \mathrm{Time}, \mathrm{TimelineElem}) \\
 & | & \mathit{procTweet}\,(\mathrm{User}, \mathrm{Time}, \mathrm{Tweet}) \\
 & | & \mathit{searchIndex}\,(\mathrm{Token}, \mathrm{Time}, \mathrm{Tweet}) \\
\mathrm{User} & = & \mathit{user}\,(\mathrm{String}) \\
\mathrm{Tweet} & = & \mathit{text}\,(\mathrm{Text}) \\
 & | & \mathit{image}\,(\mathrm{Binary}, \mathrm{Text}) \\
\mathrm{Text} & = & \mathit{plain}\,(\mathrm{String}) \\
 & | & \mathit{tokenized}\,(\mathrm{List}\,(\mathrm{Token})) \\
\mathrm{Time} & = & \mathrm{Number} \\
\mathrm{TimelineElem} & = & \mathit{tweet}\,(\mathrm{User}, \mathrm{Tweet}) \\
 & | & \mathit{follolwingYou}\,(\mathrm{User}) \\
\mathrm{Token} & = & \mathit{word}\,(\mathrm{String}) \\
 & | & \mathit{userID}\,(\mathrm{String}) \\
 & | & \mathit{hashtag}\,(\mathrm{String})
\end{array}
$$

**Figure 10** Abstract syntax grammar for $\mathscr{L}\langle$TweetLog$\rangle$

# A A Language Specification for $\mathscr{L}\langle$TweetLog$\rangle$

## A.1 Abstract Syntax

TweetLog's state is represented as a set $P$ of statements S. We use algebraic data types to define S, providing its *abstract grammar* (figure 10). A statement $\mathrm{S} \in P$ can be either a tweet (*tweeted* (User, Time, Tweet)), indicating a user tweeted some tweet at a certain time, a following relationship (*follows* (User, User, Time)), indicating that one user started following another at a certain time, a timeline entry of the form *timeline* (User, Time, TimelineElem), indicating that a certain timeline element is in one user's timeline corresponding to a certain point in time, or either a processed tweet of the form *procTweet* (User, Time, Tweet), or a search index item of the form *searchIndex* (Token, Time, Tweet), which are discussed later.

A tweet can either be textual (*text* (Text)) or image (*image* (Binary, Text)) containing the image's binary content and some caption text. Text is either a plain string or a tokenized text consisting of a list of tokens: simple words, hashtags, and user IDs.

A timeline entry is either a tweet or an indication that a user is following the timeline's owner.

## A.2 Semantics

We define the semantics for $\mathscr{L}\langle$TweetLog$\rangle$ using Natural Semantics [25] (big-step structural operational semantics). This representation is considered effective for specifying purely-declarative languages, where the evaluation order is not a part of

$$
\begin{aligned}
\mathit{tokens}\,([]) &::= \varepsilon \\
\mathit{tokens}\,(t : ts) &::= \mathit{token}\,(t)\ \text{whitespace}\ \mathit{tokens}\,(ts) \\
\text{whitespace} &::= \varepsilon \\
\text{whitespace} &::= {}'\,'\ \text{whitespace} \\
\mathit{token}\,(\text{word}\,([])) &::= \varepsilon \\
\mathit{token}\,(\text{word}\,(c : cs)) &::= \mathit{wordchar}\,(c)\ \mathit{token}\,(\mathit{word}\,(cs)) \\
\text{wordchar}\,(c) &::= c\ \mathit{where}\ c \notin \{'\,',\ '@',\ '\#'\} \\
\mathit{token}\,(\text{hashtag}\,(t)) &::= {}'\#'\ \mathit{token}\,(\mathit{word}\,(t)) \\
\mathit{token}\,(\text{userID}\,(u)) &::= {}'@'\ \mathit{token}\,(\mathit{word}\,(u))
\end{aligned}
$$

**Figure 11** BNF definition of a grammar for tokens

the semantics. We use the predicate $P \models S$ to denote that statement $S$ is true under program $P$. Trivially, a statement is true under a program if it is stated by the program:

$$\frac{S \in P}{P \models S}$$

We begin our semantic definition by defining processed tweets:

$$\frac{\begin{array}{c} P \models \mathit{tweeted}\,(U, \tau, T) \\ \mathit{replaceText}\left(T, \mathit{plain}\,(X), \mathit{tokenized}\left(X'\right), T'\right) \\ \mathit{parse}\left(\mathit{tokens}\left(X'\right), X, \varepsilon\right) \end{array}}{P \models \mathit{procTweet}\left(U, \tau, T'\right)}\ \text{PROC-TWEET} \tag{6}$$

A processed tweet exists for each (raw) tweet, such that its underlying text is tokenized. $\mathit{replaceText}\left(T, X, X', T'\right)$ is a judgment that replaces the text in a tweet according to tweet type, and is defined as follows:

$$\frac{}{\mathit{replaceText}\,(\mathit{text}\,(T), T, T', \mathit{text}\,(T'))}\ \text{REPL-TEXT}_{\text{TEXT}} \tag{7}$$

$$\frac{}{\mathit{replaceText}\,(\mathit{image}\,(B, T), T, T', \mathit{image}\,(B, T'))}\ \text{REPL-TEXT}_{\text{IMG}} \tag{8}$$

$\mathit{parse}\left(\mathit{tokens}\left(X'\right), X, \varepsilon\right)$ is a judgement that relates string $X$ with a list of its tokens: words, hashtags and user IDs. Tokenization is done using the BNF-like context-free rules in figure 11. The semantics of these rules is given in figure 12.

What we did here was define a DSL (BNF) to assist us in defining the semantics for $\mathscr{L}\langle$TweetLog$\rangle$. We call such DSLs *helper DSLs*. One more thing to note here is that although we parse a string, this part of $\mathscr{L}\langle$TweetLog$\rangle$ is not its concrete syntax, but rather a part of its semantics. This is so because parsing is a part of what TweetLog needs to do, i.e., a part of its business logic. This tokenization will later be given concrete syntax, e.g., by treating user IDs and hashtags as links.

$$\frac{}{parse\,(\varepsilon, S, S)}\,\text{PARSE}_{\text{EMPTY}} \tag{9}$$

$$\frac{parse\,(A, S, R_1) \quad parse\,(B, R_1, R)}{parse\,(AB, S, R)}\,\text{PARSE}_{\text{SEQ}} \tag{10}$$

$$\frac{}{parse\,(c, c : R, R)}\,\text{PARSE}_{\text{TOKEN}} \tag{11}$$

$$\frac{P}{parse\,(c\ where\ P, c : R, R)}\,\text{PARSE}_{\text{WHERE}} \tag{12}$$

$$\frac{T ::= T' \quad parse\left(T', S, R\right)}{parse\,(T, S, R)}\,\text{PARSE}_{\text{DEF}} \tag{13}$$

**Figure 12** The semantics of the BNF-like DSL

Now that we have a tokenized tweet, we can define a search index for tweets. Generally, a tweet can be searched by any of its tokens:

$$\frac{P \models procTweet\,(U, \tau, T) \quad replaceText\,(T, tokenized\,(X), tokenized\,(X), T) \quad \xi \in X}{P \models searchIndex\,(\xi, \tau, T)}\,\text{SEARCH}_1 \tag{14}$$

As a special case, tweets can also be found by searching the user who tweeted them:

$$\frac{P \models procTweet\,(U, \tau, T)}{P \models searchIndex\,(userID\,(U), \tau, T)}\,\text{SEARCH}_2 \tag{15}$$

Now we are ready to define the timeline. A timeline of user $U_1$ contains all tweets related to user $U_2$ if user $U_1$ follows user $U_2$.

$$\frac{P \models follows\left(U_1, user\,(U_2), \tau'\right) \quad P \models searchIndex\,(userID\,(U_2), \tau, T)}{P \models timeline\,(U_1, \tau, tweet\,(U_2, T))}\,\text{TIME-LINE}_1 \tag{16}$$

Additionally, $U_1$'s timeline also contains entries for each user $U_2$ following $U_1$:

$$\frac{P \models follows\,(U_2, U_1, \tau)}{P \models timeline\,(U_1, \tau, followingYou\,(U_2))}\,\text{TIME-LINE}_2 \tag{17}$$

# B Semantics for the Fish-Tank Model

## B.1 Abstract Syntax

A logic *term* is a *compound term*, consisting of a name and zero or more child terms, or a *number*, or a *string*, or a logic *variable*:

$$\begin{array}{lcl} term & = & name(term, \ldots, term) \\ & | & number \\ & | & string \\ & | & variable \end{array}$$

A static *goal* is an *atom*, consisting of a predicate name (*pred*) and zero or more argument terms, or a conjunction of goals (∧), or a negation of a goal (¬), or the trivial goal ⊤.

$$\begin{array}{lcl} goal & = & pred(term, \ldots, term) \\ & | & goal \wedge goal \\ & | & \neg goal \\ & | & \top \end{array}$$

A (static) *clause* consists of an atom (head) and a (static) goal (body):

$$clause \quad = \quad pred(term, \ldots, term) \vdash goal$$

A dynamic goal (*dgoal*) is an *atom*, consisting of a dynamic predicate name and zero or more argument terms, or a conjunction of dynamic goals, or a negation of a dynamic goal, or a reference to a static goal:

$$\begin{array}{lcl} dgoal & = & dpred(term, \ldots, term) \\ & | & dgoal \wedge dgoal \\ & | & \neg dgoal \\ & | & s(goal) \end{array}$$

A dynamic clause (*dclause*) consists of a (dynamic) atom as head, and a dynamic goal as body:

$$dclause \quad = \quad dpred(term, \ldots, term) \vdash dgoal$$

Finally, an *axiom* is a dynamic clause, or a *fact*, consisting of a name (*factname*) and zero or more argument terms, or a *rule,* which consisting of a fact-name with arguments as *trigger*, a goal as *guard*, and an axiom as *consequence*.

$$\begin{array}{lcl} axiom & = & dclause \\ & | & factname(term, \ldots, term) \\ & | & factname(term, \ldots, term)\{goal\} \rightarrow axiom \end{array}$$

### B.2 Static Goal Semantics

The semantics of static goal $G$ is given by the judgment $S \models G$, where $S$ is a set of static clauses. Their semantics is given by the following rules.

$\top$ is a goal that succeeds unconditionally:

$$\frac{}{S \models \top}\textsc{static}_{\top}$$

A conjunction of two goals succeeds if both goals succeed under $S$:

$$\frac{S \models G_1 \quad S \models G_2}{S \models G_1 \land G_2}\textsc{static}_{\land}$$

A negation of a goal succeed if and only if the goal fails under $S$:

$$\frac{\neg(S \models G)}{S \models \neg G}\textsc{static}_{\neg}$$

An atom succeeds if there exists a clause having that atom as head, for which the body succeeds under $S$:

$$\frac{\left(G \vdash G'\right) \in S \quad S \models G'}{S \models G}\textsc{static}_{G}$$

These rules provide big-step operational semantics, ignoring the order of execution, and we bring them here for simplicity. For execution order see Prolog's operational semantics [24].

### B.3 Axiom Semantics

The foundation of the fish-tank model is the derivation of axioms from pairs of axioms through derivation rules based on Modus Ponens. We define the relation $\langle \alpha, \beta, S \rangle \rightarrow \gamma$, denoted $\gamma$ is derived from $\alpha$ and $\beta$ under $S$, as follows:

$$\frac{\mathit{fact}(\alpha) \quad \beta = \alpha\,\{G\} \rightarrow \gamma \quad S \models G}{\langle \alpha, \beta, S \rangle \rightarrow \gamma}\textsc{derive}_{\alpha F}$$

$$\frac{\mathit{fact}(\beta) \quad \alpha = \beta\,\{G\} \rightarrow \gamma \quad S \models G}{\langle \alpha, \beta, S \rangle \rightarrow \gamma}\textsc{derive}_{\alpha R}$$

These two rules represent the cases where $\alpha$ is a fact and where $\alpha$ is a rule, respectively. The judgment $\mathit{fact}(\alpha)$ is successful if and only if $\alpha$ is a fact. The judgment $S \models G$ can have zero or more solutions (with different assignments to the variables in $G$, which may be shared with $\gamma$), yielding potentially different results.

### B.4 State of the Fish-Tank

The state of the fish-tank is represented by a tuple $(S, T, Q)$, where $S \subseteq \{clause\}$ is a set of (static) clauses ($\{clause\}$ here denotes the set of all clauses), $T \subseteq \{axiom\} \times \mathbb{Z}$ is a set of tuples $(\alpha, n)$ where $\alpha$ is an axiom and $n$ is an integer indicating how many times $\alpha$ exists in the fish-tank, and $Q \in (\{axiom\} \times \mathbb{Z})^*$ is a sequence of such $(\alpha, n)$ pairs.

Adding and removing occurrences of axioms to and from sets is done using the *add* function, defined as follows:

$$\frac{(\alpha,n) \in T}{add(T,\alpha,m) = T \setminus (\alpha,n) \cup \{(\alpha,n+m)\}}\text{ADD}_1$$

$$\frac{\forall n \in \mathbb{Z}, (\alpha,n) \notin T}{add(T,\alpha,m) = T \cup \{(\alpha,m)\}}\text{ADD}_2$$

An insertion of axiom $\alpha$ to the tank is done by appending a request to add the axiom with multiplicity 1 to $Q$:

$$\frac{}{(S,T,Q) \overset{ins(\alpha)}{\Rightarrow} (S,T,\langle Q,(\alpha,1)\rangle)}\text{INS}$$

Similarly, removal of axiom $\alpha$ from the tank is appending a request to subtract 1 from $\alpha$'s value:

$$\frac{}{(S,T,Q) \overset{rem(\alpha)}{\Rightarrow} (S,T,\langle Q,(\alpha,-1)\rangle)}\text{REM}$$

The fish-tank is updated spontaneously, removing the first element from $Q$ and applying it to $T$. During that process, derived axioms are identified and placed back to $Q$.

$$\frac{Q = \left\langle (\alpha,n), Q' \right\rangle \quad Q'' = \langle (\gamma, n \times m) \mid (\beta,m) \in seq(T) \land \langle \alpha,\beta,S\rangle \to \gamma \rangle}{(S,T,Q) \Rightarrow (S, add(T,\alpha,n), \langle Q',Q''\rangle)}\text{TICK}$$

The interesting part of the TICK rule above is the calculation of the suffix $Q''$ we add to $Q$. We calculate it using sequence comprehension. To calculate $Q''$ we first look for all elements of the form $(\beta, m)$ in $T$, where $\beta$ can interact with $\alpha$ to produce $\gamma$. The *seq* function here turns a set into a sequence, choosing an arbitrary order. Each derived axiom $\gamma$ is added with multiplicity $n \times m$, representing the amount of occurrences it would have if $n$ occurrences of $\alpha$ interacted with $m$ occurrences of $\beta$. The TICK rule is applied continuously, as long as $Q$ is not empty, updating $T$ gradually.

### B.5 Dynamic Goal Semantics

The state of the fish-tank is visible from the outside by applying queries on dynamic goals. We use the judgment $(S, T, Q) \models G^d$ to denote that dynamic goal $G^d$ succeeds under state $(S, T, Q)$.

A dynamic goal referencing a static goal succeeds if and only if the static goal succeeds under $S$:

$$\frac{S \models G}{(S,T,Q) \models s(G)}\text{QUERY}_s$$

A conjunction of dynamic goals succeeds if both goals succeed under the tank's state:

$$\frac{(S,T,Q) \models G_1^d \quad (S,T,Q) \models G_2^d}{(S,T,Q) \models G_1^d \wedge G_2^d}\text{QUERY}_\wedge$$

Here too we give big-step operational semantics, leaving the order of execution undefined for simplicity of presentation. Here too the operational semantics of Prolog provides a good reference for this.

Atomic goals depend on the existence of matching dynamic clauses in $T$. For such clauses we check that their multiplicity is positive (i.e., they actually exist in the tank) and that their body holds under the tank's state.

$$\frac{\left(G_1^d \leftarrow G_2^d, n\right) \in T \quad n > 0 \quad (S,T,Q) \models G_2^d}{(S,T,Q) \models G_1^d}\text{QUERY}_\leftarrow$$

```
1. declare tweet :: type
2. declare text :: type
3. declare user :: type
4. declare timelineElem :: type
5. declare User tweeted Tweet at Time :: axiom
     where  User :: user , Time :: number , Tweet :: tweet
     display as h  《 User :: user 》  " tweeted "  《 Tweet :: tweet 》  " at "  《 Time :: number 》
6. declare A follows B since Time :: axiom
     where  A :: user , B :: user , Time :: number
     display as h  《 A :: user 》  " follows "  《 B :: user 》  " since "  《 Time :: number 》
7. declare timeline ( User , Time , Elem ) :: pred where  User :: user , Time :: number , Elem :: timelineElem
8. declare procTweet ( User , Time , Tweet ) :: axiom where  User :: user , Time :: number , Tweet :: tweet
9. declare searchIndex ( Token , Time , User , Tweet ) :: axiom
     where  Token :: < tweetCtx > , Time :: number , User :: user , Tweet :: tweet
10. declare @ Name :: user
     where  Name :: string
     display as h  " @ "  《 Name :: string 》
11. declare " Text " :: tweet
     where  Text :: text
     display as h  " " "  《 Text :: text 》  " " "
12. declare raw Str :: text
     where  Str :: string
     display as h  1/2 " raw "  《 Str :: string 》
13. declare tokenized ( Tokens ) :: text where  Tokens :: list ( < tweetCtx > )
14. declare tweet ( User , Tweet ) :: timelineElem where  User :: user , Tweet :: tweet
15. declare User is following you :: timelineElem
     where  User :: user
     display as h  《 User :: user 》  " is following you "
16. declare w Word :: < tweetCtx >
     where  Word :: string
     display as h  1/2 " w "  《 Word :: string 》
17. declare # Tag :: < tweetCtx >
     where  Tag :: string
     display as h  " # "  《 Tag :: string 》
18. declare @ U :: < tweetCtx >
     where  U :: string
     display as h  " @ "  《 U :: string 》
```

**Figure 13** The abstract syntax of $\mathscr{L}\langle$TweetLog$\rangle$ as a collection of Cedalion type declarations

## C Code Snippets from the Implementation of TweetLog

The abstract syntax of $\mathscr{L}\langle$TweetLog$\rangle$ as a collection of Cedalion type declarations is shown in figure 13, matching the order of definitions in appendix A.1. Part of the semantics of $\mathscr{L}\langle$TweetLog$\rangle$ defining the tokenization of tweets is shown in figure 14.

```
1. token ( tweetCtx ) w Word ::= [ ^ #@ ]+! as WordCodes . { charCodes ( Word , WordCodes ) }
2. token ( tweetCtx ) @ User ::= ' @ ' . ( [ ^ ]+! as Codes ) . { charCodes ( User , Codes ) }
3. token ( tweetCtx ) # Tag ::= ' # ' . [ ^ ]+! as Codes . { charCodes ( Tag , Codes ) }
```

**Figure 14** Part of the semantics of $\mathscr{L}\langle$TweetLog$\rangle$ defining the tokenization of tweets. Based on an existing Cedalion DSL for parsing and tokenization, with semantics similar to the one depicted in figure 12.

```
:— dynamic edge/2.

go :—
    read(Cmd),
    dispatch(Cmd, Out),
    write(Out), nl,
    go.

dispatch(add_edge(X, Y), ok) :—
    assert(edge(X, Y)).
dispatch(from(X), R) :—
    findall(Y, reachable(X, Y), R).
dispatch(to(Y), R) :—
    findall(X, reachable(X, Y), R).

reachable(X, X).
reachable(X, Z) :—
    edge(X, Y),
    reachable(Y, Z).
```

**Figure 15** Prolog program that modifies itself as means of maintaining state

```
|: add_edge(a, b).
ok
|: add_edge(b, c).
ok
|: add_edge(a, e).
ok
|: to(e).
[e,a]
|: from(e).
[e]
|: from(a).
[a,b,c,e]
```

**Figure 16** Usage example of the program in figure 15

## D Program Mutation in Prolog

Prolog support program mutation as a common pattern for storing state. The Prolog program in figure 15 defines the imperative predicate go/0, which implements a Read-Evaluate-Print-Repeat (REPL) cycle, allowing users to build and query directed graphs. The command add_edge(X, Y) adds an edge to a graph from X to Y, implicitly adding vertexes X and Y if they do not exist. The commands from(X) and to(X) print vertexes with paths from or to X, respectively. In doing so their respective implementations consult the predicate reachable/2, which succeeds for all reachable node pairs in the graph. See figure 16 for example of how this program is used.

When handling add_edge(X, Y), the program uses **assert**/1 to add a fact of the form edge(X, Y) to the program, effectively modifying the program. reachable/2 is a declarative predicate that consults the mutable (dynamic) edge/2 predicate as part of its logic.

## About the authors

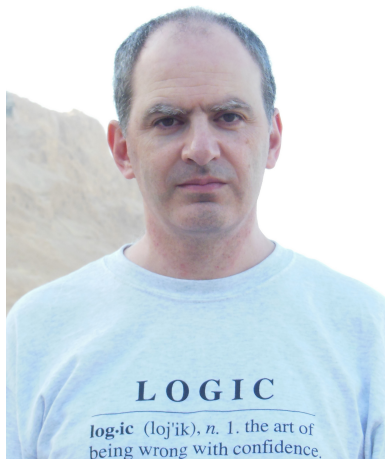

**David H. Lorenz** is an Associate Professor in the Department of Mathematics and Computer Science at the Open University of Israel. He is currently a Visiting Professor at the Faculty of Computer Science, Technion—Israel Institute of Technology. His research interests include aspect-oriented software engineering, modularity, and programming, particularly involving domain specific languages. Prof. Lorenz received his PhD in computer science from the Technion—Israel Institute of Technology. He is a member of the ACM and a member of the IEEE. Contact him at the Dept. of Mathematics and Computer Science, Open University of Israel, Raanana 4310701, Israel; lorenz@openu.ac.il.

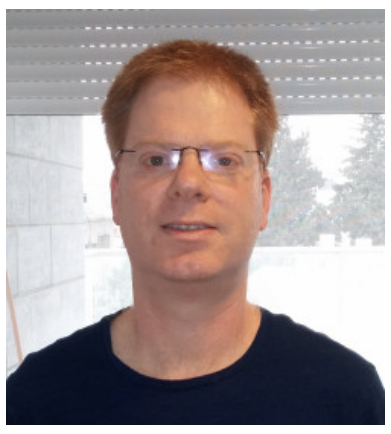

**Boaz Rosenan** is a PhD Candidate at the University of Haifa, under the supervision of Prof. David H. Lorenz. His research interests include language workbenches and domain specific languages. He received his BA in computer science from the Technion—Israel Institute of Technologies and his MSc in computer science from the Open University of Israel. He is a member of the ACM. Contact him at the Dept. Computer Science, University of Haifa, Haifa 3498838, Israel; brosenan@gmail.com.